\documentclass[aps,prl,reprint,superscriptaddress,nofootinbib,floatfix]{revtex4-2}
\usepackage{graphicx}
\usepackage{amsmath,amssymb,bm}
\usepackage{microtype}
\usepackage[colorlinks=true,citecolor=blue,linkcolor=blue,urlcolor=blue]{hyperref}
\begin{document}

\title{Spontaneous Motion Generates Reversible Nonreciprocity in an Achiral Active Elastic Ring}
\author{Wen-de Tian}
\email{tianwend@suda.edu.cn}
\affiliation{Center for Soft Condensed Matter Physics and Interdisciplinary Research  \&  School of Physical Science and Technology, Soochow University , Suzhou, 215006, P. R. China}
\author{Kang Chen}
\email{kangchen@suda.edu.cn}
\affiliation{Center for Soft Condensed Matter Physics and Interdisciplinary Research \& School of Physical Science and Technology, Soochow University , Suzhou, 215006, P. R. China}
\author{Tianhui Zhang}
\email{zhangtianhui@suda.edu.cn}
\affiliation{Center for Soft Condensed Matter Physics and Interdisciplinary Research \& School of Physical Science and Technology, Soochow University , Suzhou, 215006, P. R. China}

\begin{abstract}
Nonreciprocal mechanical response is usually associated with built-in directional couplings, structural chirality, or external driving. Whether an achiral active body can instead generate and reverse such directionality through its own motion remains less clear. Here we show that spontaneous rotation makes an active elastic ring with reciprocal passive interactions mechanically nonreciprocal. Off-center elastic forces reorient propulsion, coupling deformation back to motion. Reversing the rotation reverses the antisymmetric response, while a compensating control preserves finite directionality but strongly suppresses the near-Hopf resonance. Approaching Hopf, the selected response grows inversely with distance to onset and becomes a self-sustained traveling deformation above threshold. Thus nonreciprocity is selected by the dynamical state, while active feedback controls its critical amplification.
\end{abstract}
\maketitle

Mechanical reciprocity constrains linear response: exchanging an applied force and its conjugate displacement measurement leaves the response unchanged. For passive elasticity about a stationary equilibrium, this symmetry follows from the elastic energy. Active and driven systems can break reciprocity through nonconservative constitutive couplings, nonreciprocal interactions, or externally imposed spatiotemporal bias~\cite{Scheibner2020Odd,Scheibner2020NH,You2020,Fruchart2021,BraunsMarchetti2024,Nemeth2026,Nassar2020,Fleury2014}. Active solids, however, can also move and deform autonomously~\cite{Baconnier2022,Xu2023,HernandezLopez2024,Veenstra2025}. \emph{This raises a distinct question: can spontaneously selected motion itself generate nonreciprocal mechanical response in an achiral active body with reciprocal elasticity, and what controls the strength of that response?}

Here we show that it can in an active elastic ring. Off-center elastic forces reorient propulsion, closing a feedback loop between deformation and motion. The ring spontaneously selects one of two mirror-related rotations. Within either state, exchanging the force and displacement-measurement sites changes the response, while reversing the rotation reverses its antisymmetric part. A control that preserves the rotating state while interrupting polarity-induced changes in propulsion leaves finite directionality but suppresses the sharp near-Hopf resonance. This separates two roles of activity: the dynamical state selects the direction of mechanical response, whereas active feedback controls its critical amplification.

The same selected mode connects linear response to autonomous deformation. As activity approaches a Hopf threshold from below, the mode loses damping and its susceptibility grows inversely with the distance to onset; above threshold, it becomes a self-sustained traveling deformation. The instability persists in the continuum limit, while bending controls its nonlinear saturation, giving either continuous wave onset or coexistence with stable rotation. Thus nonreciprocal transmission, critical amplification, and autonomous deformation arise as successive stages of the same active feedback mechanism.

\begin{figure}[htbp]
\centering
\includegraphics[width=0.8\columnwidth]{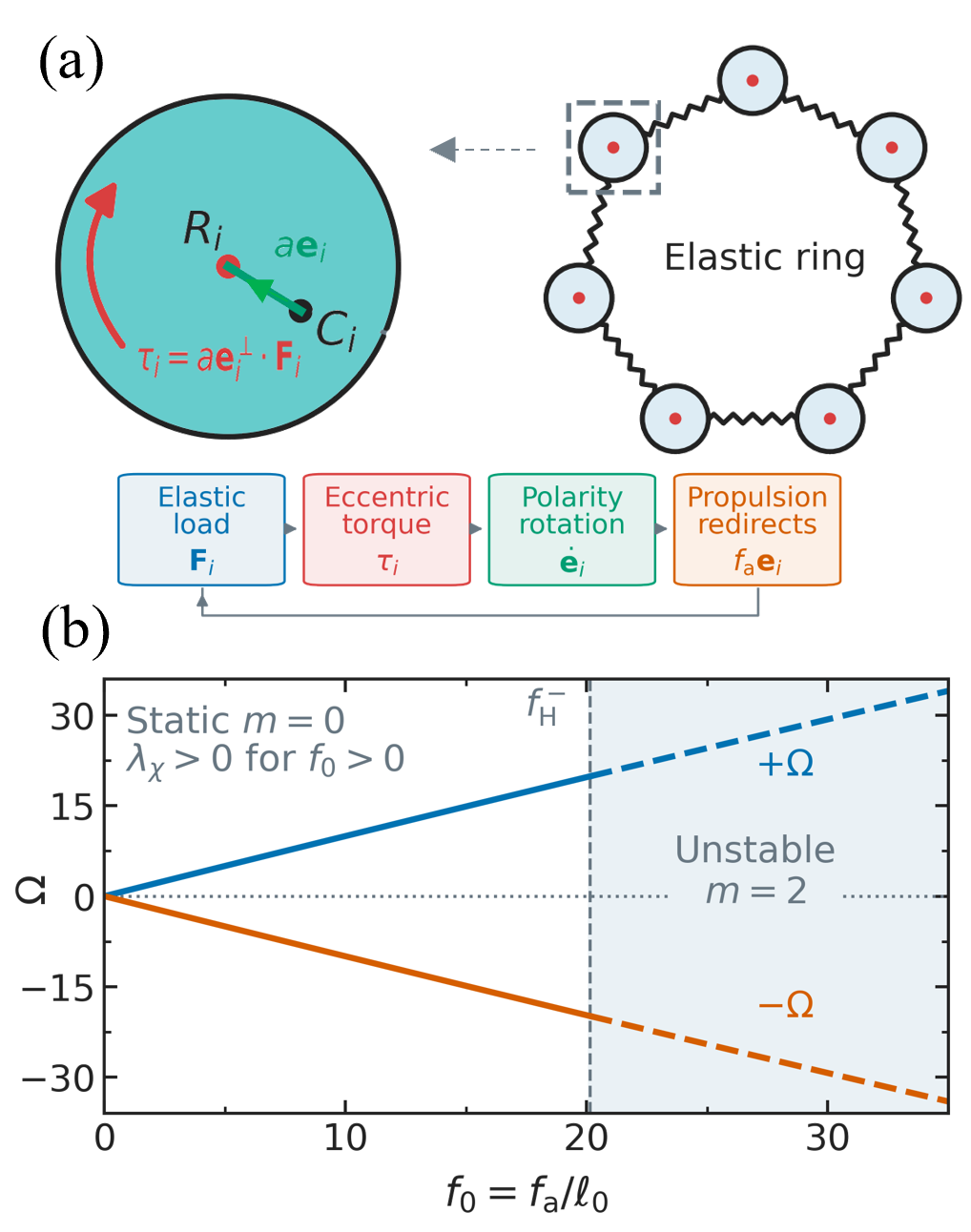}
\caption[Reciprocal elasticity and spontaneous mirror rotation.]{\textbf{Reciprocal elasticity and spontaneous mirror rotation.} \textbf{(a)} Unit geometry and feedback between elastic force, eccentric torque, polarity rotation, and propulsion. $\bm R_i$ is the drag center, $\bm C_i=\bm R_i+a\bm e_i$ the elastic connection point, and $\bm F_i$ the reciprocal elastic force; the eccentric torque is $a\bm e_i^\perp\cdot\bm F_i$. \textbf{(b)} Exact stationary and rotating solutions for $N=32$. The radial stationary state is unstable to a uniform polarity tilt, with growth rate $\lambda_\chi>0$ for $f_0>0$. Mirror-related rotations are stable against shape perturbations below the lower Hopf threshold $f_{\rm H}^-=20.137$.}
\label{fig:model}
\end{figure}

\emph{Microscopic model and force--torque feedback.}---
We consider a two-dimensional ring of $N$ active units. Unit $i$ has a
translational drag center $\bm R_i$, an in-plane polarity
$\bm e_i=(\cos\theta_i,\sin\theta_i)$, and an elastic connection point
$\bm C_i=\bm R_i+a\bm e_i$, where the fixed offset $a$ is the
eccentricity [Fig.~\ref{fig:model}(a)]. Neighboring connection points are
coupled by reciprocal stretching and bending interactions. The passive energy
is $U=U_{\rm s}+U_{\rm b}+U_{\rm WCA}$, with
$U_{\rm s}=(k_\ell/2)\sum_i(|\bm C_{i+1}-\bm C_i|-\ell_0)^2$ and
$U_{\rm b}=\kappa_\theta\sum_i[1-\cos(\alpha_i-2\pi/N)]$.
Here $k_\ell$ is the stretching stiffness, $\ell_0$ the bond rest length,
$\kappa_\theta$ the bending-energy scale, and $\alpha_i$ the signed turning
angle between adjacent bonds. The short-range Weeks--Chandler--Andersen
(WCA) potential $U_{\rm WCA}$ prevents nonbonded overlap but is inactive in the
states used below. All passive forces derive from this reciprocal
potential, with $\bm F_i=-\partial U/\partial\bm C_i$.

In the overdamped limit, the deterministic dynamics is
\begin{equation}
\zeta_{\rm t}\dot{\bm R}_i=f_{\rm a}\bm e_i+\bm F_i,\qquad
\zeta_{\rm r}\dot\theta_i=a\,\bm e_i^\perp\!\cdot\bm F_i ,
\label{eq:micro}
\end{equation}
where overdots denote time derivatives, $f_{\rm a}$ is the self-propulsion force magnitude, $\zeta_{\rm t}$ and $\zeta_{\rm r}$
are the translational and rotational drag coefficients, and
$\bm e_i^\perp=(-\sin\theta_i,\cos\theta_i)$. In two dimensions, eccentricity does not
introduce a nonreciprocal interaction; it provides a purely geometric
force--torque conversion.

Eliminating $\bm F_i$ gives the relation $\dot\theta_i=[a/(a^2+\ell_{\rm r}^2)]\bm e_i^\perp\!\cdot\dot{\bm C}_i$, with $\ell_{\rm r}=\sqrt{\zeta_{\rm r}/\zeta_{\rm t}}$. Transverse connection-point motion rotates the polarity, closing the load--torque--orientation--propulsion loop [Fig.~\ref{fig:model}(a)]. We vary activity through the propulsion force per reference length, $f_0=f_{\rm a}/\ell_0$. The model parameters are the stretching and bending moduli, $K_{\rm s}=k_\ell\ell_0$ and $K_{\rm b}=\kappa_\theta\ell_0$, and drag coefficients per reference length, $\xi_{\rm t,r}=\zeta_{\rm t,r}/\ell_0$. We set  $N=32$, $R_0=1$, $K_{\rm s}=1.266$, $K_{\rm b}^\ast=3.004\times10^{-4}$, $\xi_{\rm t}=1$, $\xi_{\rm r}=1.533\times10^{-4}$, and $a=0.119$. Here $R_0$ is the reference ring radius, as the length unit; $K_{\rm b}^\ast$ denotes the fixed bending modulus. Only $f_0$ varies unless otherwise stated.The particle-parameter mapping is given in the End Matter. Robustness test of eccentricity and noise are given in Supplemental Material (SM)\cite{SM}.

In the stationary state, the connection points form a regular polygon, and each polarity points radially outward~\cite{SM}. Its propulsion is balanced by elastic tension, the state cannot undergo a Hopf bifurcation. A uniform polarity tilt destabilizes this stationary state at $f_0>0$. Opposite tilts lead to mirror-related rotations $\pm\Omega$ with otherwise identical properties [Fig.~\ref{fig:model}(b)], as confirmed by independent particle simulations~\cite{SM}. This means that the initially achiral ring  acquires chirality by spontaneously choosing a rotation direction.

\begin{figure}[htbp]
\centering
\includegraphics[width=0.8\columnwidth]{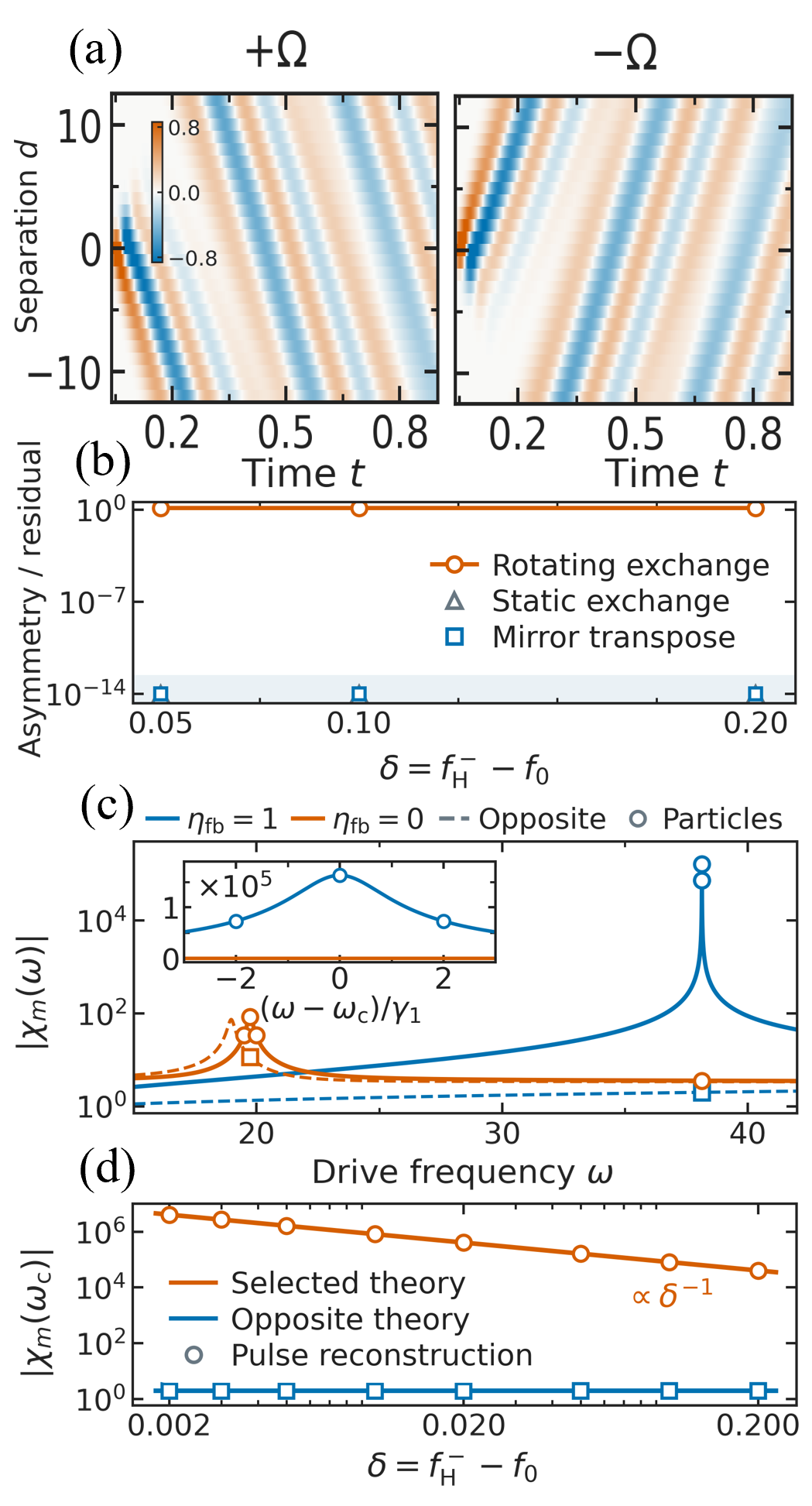}
\caption[Reversible directional response and feedback-controlled amplification.]{\textbf{Reversible directional response and feedback-controlled amplification.} \textbf{(a)} Particle pulse responses for $\pm\Omega$, at $\delta=f_{\rm H}^--f_0=0.050$. \textbf{(b)} Exchange asymmetry, $\|\mathbf H-\mathbf H^{\mathsf T}\|_{\rm F}/\|\mathbf H\|_{\rm F}$, for rotating and stationary states, and normalized error in the mirror-transpose relation; $\|\cdot\|_{\rm F}$ is the Frobenius norm.  \textbf{(c)} Original ($\eta_{\rm fb}=1$) and compensated ($\eta_{\rm fb}=0$) frequency responses.  Inset: narrow original resonance in $10^5$ susceptibility units.   \textbf{(d)} Selected/opposite traveling-polarization susceptibilities at the same positive frequency for each $\delta$: theory (lines) and particle pulse measurements (symbols).}
\label{fig:response}
\end{figure}

\emph{Motion-generated nonreciprocity.}---
When the rotating ring is stable against shape perturbations, we apply a weak radial force at connection point $\bm C_j$ and measure radial displacement at $\bm C_i$. The source $j$ and receiver $i$ are fixed particle labels. The complex response $\mathbf H_{\Omega,ij}(\omega)$ gives the displacement amplitude per unit force and its phase relative to the force at angular frequency $\omega$ (End Matter).The applied force also produces the eccentric torque.

A localized force pulse produces different displacement responses at positive and negative separations along the ring. The preferred propagation direction reverses between $\pm\Omega$ [Fig.~\ref{fig:response}(a),Movie S1~\cite{SM}], before any self-sustained shape wave appears. A small displacement changes the reciprocal elastic force. The force rotates the polarity, whose redirected propulsion changes the subsequent deformation.  In the rotating state, the polarity tilt and rotation break this symmetry, so perturbations traveling in opposite directions experience different feedback amplitudes and phases. It should be noticed that measuring deformation propagation within the ring alone cannot determine how much of the response amplification arises from active feedback and how much from other dynamical effects of the rotating state. 

Exchanging the force-source and displacement-measurement positions directly tests reciprocity, which requires $\mathbf H_\Omega=\mathbf H_\Omega^{\mathsf T}$. Using the same radial forcing and displacement measurements, we compare the stationary and rotating states (two distinct active states of the same paramters with reciprocal passive elasticity). We find the force-balanced stationary reference has a reciprocal radial response, whereas the stable rotating state does not [Fig.~\ref{fig:response}(b)]. Additionally, for the regular ring, mirror symmetry gives $\mathbf H_{-\Omega}(\omega)=\mathbf H_{+\Omega}^{\mathsf T}(\omega)$. Reversing rotation changes the sign of the antisymmetric response and preserves the symmetric response. The same reversal appears in the displacement response to a short force pulse. This implies that the preferred transmission direction reverses between mirror-related rotating states without changing  parameters. 

To isolate the amplification mechanism, we add a compensating force at each drag center,
$\bm F_i^{\rm servo}=-(1-\eta_{\rm fb})f_{\rm a}[\bm e_i(t)-\bm e_i^{(0)}(t)]$.
The reference polarity $\bm e_i^{(0)}(t)$ follows the exact unperturbed rotating solution~\cite{SM}.  At $\eta_{\rm fb}=0$, compensation cancels the change in translational propulsion caused by polarity deviations from the reference, while retaining the eccentric elastic torque. This force could use an externally maintained rotating reference to interrupt the influence of polarity perturbations on translational propulsion. We find the compensating force suppresses the narrow resonance, while a finite directional response survives at a different frequency. Figure~\ref{fig:response}(c) compares the original ($\eta_{\rm fb}=1$) and compensated ($\eta_{\rm fb}=0$) frequency responses on the same axes.  Direct harmonic particle forcing at $f_0=f_{\rm H}^--0.050$ gives susceptibility magnitude $|\chi_{\rm sel}|=1.628\times10^5$ in the original system and $84.360$ at the compensated system's shifted resonance. Within each system, the selected/opposite susceptibility ratio at the same positive frequency is $8.170\times10^4$ and $7.290$, respectively. At the original resonance the compensated response is only $3.550$.  Here we call the traveling polarization resonant at positive frequency in the original system ``selected''; the ``opposite'' polarization represents a deformation traveling in the opposite direction. 

In the original system, the mode fof a twofold deformation of the ring has eigenvalue $\lambda_{\rm c}=-\gamma+\mathrm{i}\omega_{\rm c}$, where $\gamma$ is its decay rate and $\omega_{\rm c}$ its oscillation frequency. 
As the activity approaches the Hopf threshold from below, the  mode decays more slowly. A weak periodic force applied at the mode’s oscillation frequency produces a larger deformation. The response grows inversely with the difference between the Hopf threshold and the activity, $|\chi_{\rm sel}(\omega_{\rm c})|\propto\gamma^{-1}\propto\delta^{-1}$ [Fig.~\ref{fig:response}(d)], while the response for the opposite propagation direction remains finite at the same frequency. It means that changes in polarity redirect propulsion and strongly amplify the resonant response. Directional asymmetry persists when this effect is cancelled, showing that this feedback is not the sole source of nonreciprocity. It is because the compensation leaves the coupling between radial and tangential deformation on the rotating ring intact. This coupling produces a finite directional response even when polarity changes no longer redirect propulsion. The latter feedback strongly enhances the near-Hopf resonance by reducing damping and increasing the coupling between the force and the measured displacement~\cite{SM}. Moreover, responses reconstructed from short force pulses agree with measurements under sustained periodic forcing and with the full microscopic linear-response theory~\cite{SM}.  

\begin{figure}[htbp]
\centering
\includegraphics[width=0.8\columnwidth]{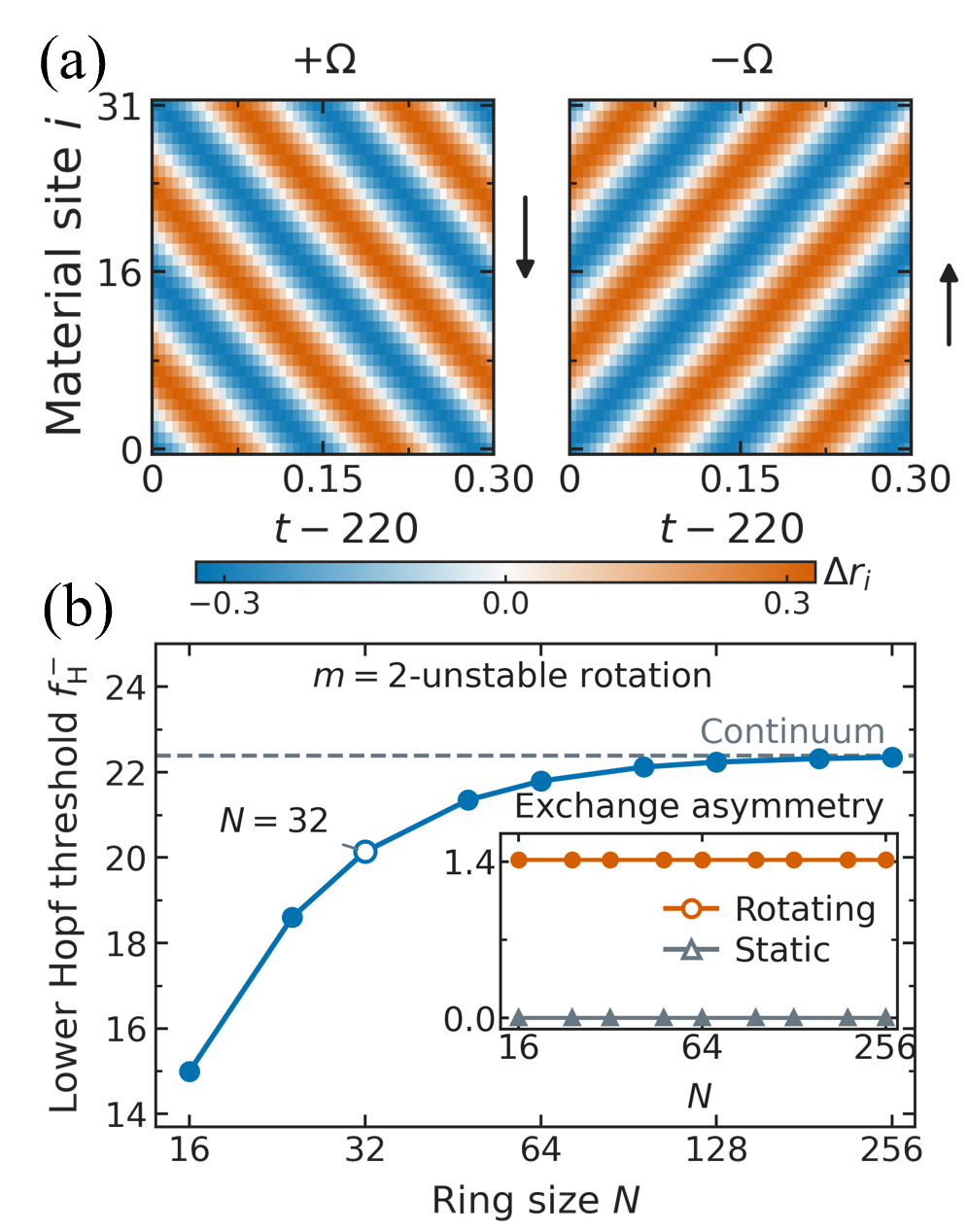}
\caption[Traveling deformation and the continuum limit.]{\textbf{Traveling deformation and the continuum limit.} \textbf{(a)}  Radial deformation $\Delta r_i$ relative to the site-averaged radius about the connection centroid.  \textbf{(b)} Lower Hopf thresholds converge toward the independently calculated continuum value (dashed). Inset: exchange asymmetry for rotating and stationary states as $N$ increases, with the vertical axis starting at zero. }
\label{fig:finite}
\end{figure}

\emph{From critical amplification to a self-sustained traveling wave.}---
As activity approaches the Hopf threshold, the response is  forced to grow. Above threshold, the deformation grows spontaneously and reaches a finite amplitude, forming a traveling wave sustained by active feedback.

The two deformation maxima pass through successive particles, showing propagation relative to the material [Fig.~\ref{fig:finite}(a), Movie S2~\cite{SM}]. A rigidly rotating ellipse would keep its maxima on the same particle labels. Reversing rotation reverses wave propagation because it interchanges how oppositely traveling perturbations couple to the tilted polarities. Near onset, particle simulations reproduce the growth rates predicted by the full microscopic linear theory, while numerical continuation connects the wave branch to the Hopf point~\cite{SM}.   It implies that the amplified response below threshold and the autonomous wave above it arise from the same mode.  When $N$ approches infinite, the lower Hopf threshold approaches $f_{\rm H}^-(\infty)=22.384$ [Fig.~\ref{fig:finite}(b)], while the normalized radial exchange asymmetry remains finite. Besides, the continuum stability theory and the response calculations also support the persistence of oscillatory onset and nonreciprocal transmission, respectively (see End Matter and SM~\cite{SM}).

\begin{figure}[htbp]
\centering
\includegraphics[width=0.75\columnwidth]{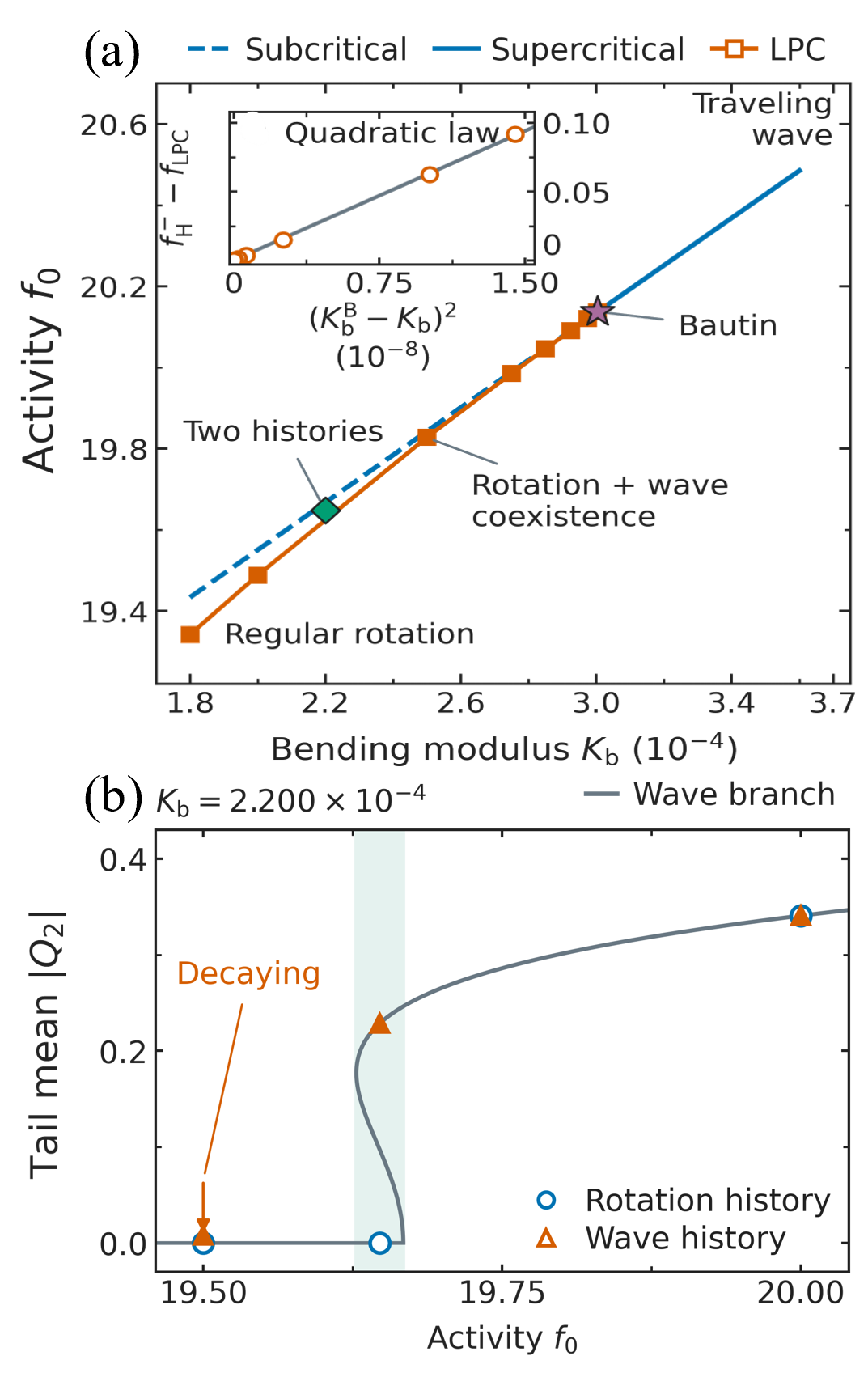}
\caption[Bending controls wave onset and coexistence.]{\textbf{Bending controls wave onset and coexistence.} \textbf{(a)} Hopf and limit-point-of-cycles (LPC) boundaries at $N=32$ meet at the generalized-Hopf (Bautin) point (star). Shading marks coexistence of stable rotation and a stable traveling wave; the diamond marks simulations with two initial conditions at identical parameters.  Inset: coexistence widths  follow the quadratic law. \textbf{(b)} Wave amplitudes from continuity theory (gray) and particle amplitudes averaged over the final observation window, starting near rotation (circles) or a wave (triangles), at $K_{\rm b}=2.200\times10^{-4}$. Within the coexistence region the two initial conditions lead to different long-time states; above Hopf they approach the same wave. Below LPC the wave-seeded run is still decaying (arrow).}
\label{fig:nonlinear}
\end{figure}

\emph{Bending controls wave onset and coexistence.}---
The Hopf boundary marks where an oscillating shape perturbation begins to grow. Bending also controls how this growth reaches a finite amplitude. Near the Hopf onset, the deformation is dominated by the critical $m=2$ traveling mode. Let $A$ denote its amplitude, with $|A|\propto |Q_2|$ near threshold (End Matter). Its slow dynamics obey
$\partial_t |A| \simeq \mu |A| +c_3 |A|^3 +c_5 |A|^5 $, where $\mu$ is the linear growth rate. This normal form explains how bending changes the nature of the wave onset: $c_3<0$ gives a continuous transition, whereas $c_3>0$, together with $c_5<0$, allows a finite-amplitude traveling wave to coexist with stable rotation. The cubic coefficient changes sign at the Bautin point, whose bending modulus is $K_b^{\rm B}$ [Fig.~\ref{fig:nonlinear}(a)]. For $K_{\rm b}>K_{\rm b}^{\rm B}$, waves emerge continuously from zero amplitude at Hopf. For $K_{\rm b}<K_{\rm b}^{\rm B}$, finite-amplitude waves persist below Hopf, where shape-preserving rotation remains stable. The stable and unstable wave branches meet at a limit point of cycles (LPC). Between this boundary and Hopf, $f_{\rm LPC}<f_0<f_{\rm H}^-$, the initial condition determines which state is reached. Near the Bautin point, the coexistence interval widens according to $f_{\rm H}^--f_{\rm LPC}\propto(K_{\rm b}^{\rm B}-K_{\rm b})^2$ [inset of Fig.~\ref{fig:nonlinear}(a)], consistent with the Bautin normal form~\cite{Kuznetsov2023,SM}. 

Particle simulations indeed demonstrate this dependence on initial conditions [Fig.~\ref{fig:nonlinear}(b)]. At $(K_{\rm b},f_0)=(2.200\times10^{-4},19.648)$, runs started near rotation and near a traveling wave approach different long-time states. The wave has a mean twofold deformation $|Q_2|\simeq0.2286$ over the final observation window (End Matter; Movie S3~\cite{SM}). Below LPC, the wave decays; above Hopf, both initial conditions approach the same wave. Changing bending stiffness alters both the resistance to deformation and the elastic forces that turn the polarities, thereby changing the instability threshold and whether a finite-amplitude wave can coexist with stable rotation.

\emph{Discussion and outlook.}---
We highlight that nonreciprocity need not be built into the passive interactions of an active material. Here an achiral system spontaneously chooses a rotating state, and that dynamical choice selects the sign of its mechanical directionality. Reversing the rotation reverses the antisymmetric response without changing parameters. The compensation experiment further separates two roles of activity: motion selects the preferred transmission direction, whereas polarity--propulsion feedback controls the large near-critical gain.

This state-selected response also links sensing and self-organization. The directional mode amplified below Hopf becomes the traveling deformation above it, so linear amplification and autonomous wave propagation are two regimes of the same active mode. Our results therefore suggest a route to mechanically reconfigurable active connected matterials in which motion itself writes, reverses, and amplifies directional response. How generally such motion-written nonreciprocity emerges beyond the symmetric ring remains an open question.

\emph{Acknowledgments:} This work was supported by the National Natural Science Foundation of China (Grant No. 22573069).  K. Chen acknowledges financial support from NSFC Grants No. 21774091 and T. Zhang acknowledges financial support from NSFC Grants No. 12574230
OpenAI was used for language polishing. 

\emph{Data Availability:} They are available from the corresponding author upon reasonable request.

\section*{End Matter}
\emph{Model parameters and continuum theory.}---
The  parameters used in Figs.~1--3 are $R_0=1$, $K_{\rm s}=1.266$, $K_{\rm b}^\ast=3.004\times10^{-4}$, $\xi_{\rm t}=1$, $\xi_{\rm r}=1.533\times10^{-4}$, and $a=0.119$. Here $K_{\rm s}=k_\ell\ell_0$ and $K_{\rm b}=\kappa_\theta\ell_0$ are material moduli, while $\xi_{\rm t,r}=\zeta_{\rm t,r}/\ell_0$ are drag densities. Lengths are measured in units of the reference radius $R_0$. For each $N$, set $\ell_0=2R_0\sin(\pi/N)$ and reconstruct $k_\ell=K_{\rm s}/\ell_0$, $\kappa_\theta=K_{\rm b}/\ell_0$, $\zeta_{\rm t,r}=\xi_{\rm t,r}\ell_0$, and $f_{\rm a}=f_0\ell_0$. When $N$ increases, $K_{\rm s}$, $K_{\rm b}^\ast$, $\xi_{\rm t,r}$, $a$, and $R_0$ remain fixed; particle parameters change through $\ell_0(N)$. Figure~4 instead varies $K_{\rm b}$ and $f_0$ at $N=32$, with the other parameters unchanged. 

The continuum fields are the connection curve $\bm C(s,t)$ and the independent polarity $\bm e(s,t)=(\cos\theta,\sin\theta)$, where $s$ is ring arclength and $\theta$ is a laboratory-frame angle. The equations are
\begin{equation}
\begin{aligned}
\xi_{\rm t}(\partial_t\bm C-a\partial_t\theta\,\bm e^\perp)&=f_0\bm e+\bm f_{\rm el},\\
\xi_{\rm r}\partial_t\theta&=a\bm e^\perp\!\cdot\bm f_{\rm el},
\end{aligned}
\label{eq:continuum}
\end{equation}
with $\bm f_{\rm el}$ the force per unit reference arclength derived from reciprocal stretching and bending energy~\cite{SM}. The nonzero continuum Hopf threshold is obtained by linearizing these equations about the rotating state. Persistence of radial exchange asymmetry is checked separately by calculating force--displacement responses at increasing $N$ with the same material parameters. The stability calculation and response calculation test oscillatory onset and directional transmission, respectively.

\emph{Conjugate transfer and mirror reversal.}---
Let $\mathbf J_\Omega$ be the Jacobian governing small particle perturbations in the frame rotating at $\Omega$. The matrix $\mathbf D$ converts particle-position and polarity perturbations into radial connection-point displacements, and $\mathbf M$ is the diagonal matrix of translational and rotational mobilities. The work of a radial force at $\bm C_j$ is conjugate to the radial displacement of that same connection point, so the force enters the dynamics through $\mathbf B=\mathbf M\mathbf D^{\mathsf T}$:
\begin{equation}
\mathbf H_\Omega(\omega)=\mathbf D[\mathrm{i}\omega\mathbf I_{3N}-\mathbf J_\Omega]^{-1}\mathbf M\mathbf D^{\mathsf T}.
\end{equation}
The inverse matrix $[\mathrm{i}\omega\mathbf I_{3N}-\mathbf J_\Omega]^{-1}$ is the resolvent, which gives the linear response of all particle variables at frequency $\omega$. The magnitude and phase of each element of $\mathbf H_\Omega$ give the displacement amplitude per unit force and its phase relative to the force. Omitting the torque would instead probe at the drag center and break conjugacy with the connection-point readout. The stationary reference and rotating state use the same $N$, $f_0$, $K_{\rm b}$, other material parameters, $\omega$, and radial force--displacement measurements, but their force-balanced geometries and elastic tensions differ. At the three tested detunings the static exchange asymmetry is below $6\times10^{-15}$, whereas the rotating one approaches $\sqrt2$.

For the regular ring, rotational symmetry makes the response depend only on particle-label separation (a circulant response matrix). With the same labels for both rotation directions, the mirror relation is $\mathbf H_{-\Omega}=\mathbf H_{+\Omega}^{\mathsf T}$. With $\mathbf H^{\rm S}=(\mathbf H+\mathbf H^{\mathsf T})/2$ and $\mathbf H^{\rm A}=(\mathbf H-\mathbf H^{\mathsf T})/2$, reversal preserves $\mathbf H^{\rm S}$ and changes the sign of $\mathbf H^{\rm A}$. Reflecting the particle labels gives an equivalent form of the same mirror relation; that relabeling must not be applied again after the transpose. Normalized Fourier vectors $(\bm\varphi_m)_j=N^{-1/2}e^{2\pi\mathrm{i}mj/N}$ define $\chi_m=\bm\varphi_m^\dagger\mathbf H\bm\varphi_m$; these susceptibilities measure responses to spatial Fourier force patterns and complement the test that exchanges a localized force source and displacement receiver. Expressing the same motion in a different coordinate frame does not by itself remove its effects on mechanical transmission~\cite{SM}.

\emph{Pole decomposition of the controlled spectra.}---
At fixed $N=32$, $f_0=20.087$, changing $\eta_{\rm fb}$ from $1$ to $0$ leaves the exact rotating state unchanged. It changes only the Jacobian terms coupling polarity perturbations to translational velocity by $-(1-\eta_{\rm fb})(f_{\rm a}/\zeta_{\rm t})\bm e^{(0)\perp}$; the elastic Hessian, rotational equation, source, and readout are retained. The compensating force requires an externally maintained rotating reference. It does not turn the system into a passive material or cancel all feedback terms. The work done by this force is included in the numerical power balance~\cite{SM}.

Near an isolated resonance of the selected deformation mode,
\begin{equation}
\chi_{\rm sel}(\omega)=\frac{\mathcal R_{\rm c}}{\gamma+\mathrm{i}(\omega-\omega_{\rm c})}+\chi_{\rm reg}(\omega),
\end{equation}
where $\mathcal R_{\rm c}=\mathcal O_{\rm in}\mathcal O_{\rm out}$ is the pole residue: $\mathcal O_{\rm in}$ measures excitation of the mode by the applied force, and $\mathcal O_{\rm out}$ measures its contribution to the observed displacement. These factors use left and right eigenvectors normalized to have unit mutual overlap. The original system has $(\gamma,\omega_{\rm c},|\mathcal R_{\rm c}|)=(9.486\times10^{-4},38.131,154.518)$; compensation gives $(0.108,19.737,8.747)$. Thus the original damping is about 114 times smaller and its residue about 18 times larger. These are analytic pole decompositions, independently tested by the particle response, not separately measured microscopic work contributions. The nonresonant term $\chi_{\rm reg}(\omega)$ is retained in the plotted spectra. Peak-response ratios between the original and compensated systems compare each near its own resonance. Within either system, selected and opposite polarizations are always compared at the same positive frequency. Stability has not been established for intermediate $0<\eta_{\rm fb}<1$.

\emph{Wave amplitude and nonlinear onset.}---
The measured second shape moment, which quantifies the twofold deformation of the connection polygon, is $Q_2=N^{-1}\sum_i(z_i-z_{\rm cm})^2$, with $z_i=C_{ix}+\mathrm{i}C_{iy}$ and $z_{\rm cm}=N^{-1}\sum_i z_i$. Near Hopf, $|A|$ is proportional to $|Q_2|$. The phase of $Q_2$ measured in the laboratory frame includes both rotation of the whole ring and propagation relative to particle labels; $\omega_{\rm int}=2\Omega-d\arg Q_2/dt$ vanishes when an elliptical deformation stays fixed relative to particle labels while the ring rotates. A traveling wave has a nonzero phase rate relative to those labels.

Numerical continuation of periodic solutions determines the sign of the effective cubic coefficient and the LPC where stable and unstable cycles meet. Supplemental Material gives the square-root amplitude scaling near LPC, the quadratic coexistence-width law, and the procedure for estimating amplitude-equation coefficients~\cite{SM}.

\end{document}